\documentclass[reprint,
    aps,prapplied,
    superscriptaddress,
    longbibliography,
    balancelastpage
]{revtex4-2}

\usepackage{graphicx}
\usepackage{siunitx}
\usepackage{amssymb,amsmath}    
\usepackage{mathptmx}
\usepackage{ORCIDinREVTeX}

\usepackage[unicode=true,
    colorlinks=true,
    linkcolor=blue,         
    citecolor=blue,         
    urlcolor=blue]{hyperref}
    
\makeatletter
\newcommand\figref[1]{%
    \@ifnextchar\bgroup{\figref@double{#1}}{\figref@single{#1}}%
}
\newcommand\figref@single[1]{\hyperref[{#1}]{\textup{FIG.~\ref*{#1}}}~}
\newcommand\figref@double[2]{\hyperref[{#1}]{\textup{FIG.~\ref*{#1}}\textcolor{blue}{{#2}}}}
\makeatother
\renewcommand{\eqref}[1]{\hyperref[{#1}]{\textup{(\ref*{#1}})}}
\newcommand{\secref}[1]{\hyperref[{#1}]{\textup{Sec.~\ref*{#1}}}}
\newcommand{\tabref}[1]{\hyperref[{#1}]{\textup{TABLE~\ref*{#1}}}}
\newcommand{\womega}{\widetilde{\omega}}
\newcommand{\bomega}{\overline{\omega}}
\newcommand{\vect}[1]{\mathbf{#1}}          

\begin{document} 

\title{Communication-Free Robust Wireless Power Transfer with Constant Output Power and \\ Stable Frequency}

\author{Zhuoyu Zhang}
\orcid{0009-0002-8020-9770}
\thanks{These authors contributed equally to this work.}
\affiliation{\mbox{School of Electrical Engineering, Xi'an Jiaotong University, Xi'an 710049, China}}%

\author{Junan Lai}
\orcid{0009-0001-4990-7354}
\thanks{These authors contributed equally to this work.}
\affiliation{\mbox{School of Electrical Engineering, Xi'an Jiaotong University, Xi'an 710049, China}}%

\author{Yuangen Huang}
\orcid{0000-0002-2486-2778}
\thanks{These authors contributed equally to this work.}
\affiliation{\mbox{School of Electrical Engineering, Xi'an Jiaotong University, Xi'an 710049, China}}%

\author{Xianglin Hao}
\orcid{0000-0001-7149-0113}
\affiliation{\mbox{Department of Electrical Engineering, City University of Hong Kong, Hong Kong 999077, China}}%

\author{Ke Yin}
\orcid{0000-0002-8534-216X}
\affiliation{\mbox{School of Electronics and Information Engineering, Sichuan University, Chengdu 610101, China}}%

\author{Zhiqin Jiang}
\orcid{0009-0005-7159-1156}
\affiliation{\mbox{School of Electrical Engineering, Xi'an Jiaotong University, Xi'an 710049, China}}%
    
\author{Chao Wang}
\orcid{0009-0004-2173-7252}
\affiliation{\mbox{School of Electrical Engineering, Xi'an Jiaotong University, Xi'an 710049, China}}%
    
\author{Xikui Ma}
\orcid{0000-0003-4138-7443} 
\affiliation{\mbox{School of Electrical Engineering, Xi'an Jiaotong University, Xi'an 710049, China}}%
    
\author{Ming Huang}
\orcid{0000-0002-3699-3468}
\affiliation{\mbox{Department of Electrical Engineering, Northwestern Polytechnical University, Xi'an 710072, China}}%

\author{Tianyu Dong}
\orcid{0000-0003-4816-0073}
\email[Corresponding author. Email: ]{tydong@mail.xjtu.edu.cn}
\affiliation{\mbox{School of Electrical Engineering, Xi'an Jiaotong University, Xi'an 710049, China}}%

\date{August 28, 2024}

\begin{abstract}
A primary challenge in wireless power transfer (WPT) systems is to achieve efficient and stable power transmission without complex control strategies when load conditions change dynamically. Addressing this issue, we propose a third-order pseudo-Hermitian WPT system whose output characteristics exhibit a stable frequency and constant power. The frequency selection mechanism and energy efficiency of the nonlinear WPT system based on pseudo-Hermitian under the coupling mode theory approximation are analyzed. Theoretical analysis indicates that under certain coupling coefficients and load conditions, the proposed system can achieve frequency adaptation in a stable frequency mode without the need to change the circuit frequency. When the load changes dynamically, the stability of the power output is maintained using a proportional integral (PI) control strategy that only collects the voltage and current at the transmitting end, eliminating the need for wireless communication circuits with feedback from the receiving side. Experimental results demonstrate that the proposed design scheme can achieve constant power transmission when load conditions change, maintaining stable and relatively high transmission efficiency. The proposed scheme exhibits benefits in practical applications since no communication is required.
\end{abstract}


\maketitle   

\section{Introduction} \label{sec:introduction}
Wireless power transfer (WPT) is one of the most intriguing innovations in modern technologies, which has received widespread attention and extensive research efforts in recent years \cite{zhang2018wireless,hui2013critical}. Using high-frequency magnetic fields, WPT systems enable wireless transmission of electricity from a power source to a designated load \cite{sample2010analysis}, showing fascinating applications in various fields \cite{patil2017wireless,buja2015design}, including implantable medical devices, electronic charging, and dynamic charging of electric vehicles \cite{lu2015wireless,dai2017safe,liu2017lab}. However, conventional WPT systems show sensitivity to variations in system parameters, with deviations in operating frequency from resonance frequency, resulting in a reduction in transmission efficiency and output power \cite{kurs2007wireless}, frequency bifurcation, impedance alteration, insufficient output power, etc. Addressing these issues requires a significant breakthrough in both theoretical understanding and technological advancements. The concept of parity-time (PT) symmetry, originating in quantum physics, has provided novel insights into the intricacies of wireless power transmission mechanisms and the architectural design of such systems \cite{schindler2011experimental}.

In 2017, a PT-symmetric WPT system was proposed, in which the energy stored in the transmitter and receiver resonators remains equitably distributed within the PT-symmetric phase, ensuring high transmission efficiency despite variations in the coupling coefficient \cite{assawaworrarit2017robust}. However, the output power is limited to $19.7~\si{mW}$ because the nonlinear gain is achieved by operational amplifiers, adequate only to power a single LED. In recent years, significant progress has been witnessed in PT symmetry theory, ranging from second-order to higher-order systems, with the aim of extending transmission distances or achieving multi-load power delivery \cite{sakhdari2020robust,zeng2020high,kim2022wide,wu2022generalized,hao2023frequency}. In addition, high-efficient wireless power transfer can be achieved using the interleaving of the gain spectrum resulting from dispersion \cite{hao2024dispersive}. Currently, driven by the emergence of power electronics technology, the methodologies to implement negative resistance at the transmitting end have evolved from a single operational amplifier to encompass comprehensive solutions such as full-bridge and half-bridge inverters \cite{zhou2018nonlinear,zhang2020omnidirectional}, Class-E power amplifiers \cite{assawaworrarit2020robust}, and other innovative approaches. These notable expansions have significantly increased the output power of PT-symmetric circuits and have found practical applications in various domains, including the charging of household appliances and the provision of power for implantable medical devices \cite{wu2021robust,wu2023robust,rong2021wireless}.

In practical applications, it is typically required that the output power at the system's receiving end remain stable. In traditional PT-symmetric systems, the output power in the symmetric region is positively correlated with the load resistance at the receiving end. Therefore, any variation in the load resistance will result in a corresponding change in the system's output power. To address this issue, a negative feedback control approach based on Buck circuit voltage regulation and closed-loop voltage control has been proposed, achieving constant power transmission independent of load resistance within a certain range \cite{zhu2020output,rong2021wireless}. However, because of the spectral bifurcation issue in second-order PT systems, tracking different frequencies in the symmetric region undoubtedly complicates the entire system. In particular, when the self-resonant frequency of the LC circuit is elevated (on the order of MHz), the system exhibits a substantial range of frequency variation, thereby imposing constraints on the selection of components. A three-coil PT-symmetric WPT system is capable of maintaining stable power and efficiency output characteristics when the total transmission distance varies, while maintaining the PT symmetry condition is a challenge, limiting the practicality of the three-coil PT symmetric WPT system. Sakhdari et al. purposely propose a method to adjust the angle between the transmitter coil and the relay coil to change the mutual inductance between the two to satisfy the PT symmetry condition, while this method requires additional spatial degrees of freedom and would increase the complexity of the design of the spatial structure of the system \cite{sakhdari2020robust}.

To address the aforementioned issues, a third-order stable frequency non-Hermitian WPT system with constant output power is proposed, which enables response to load variations without auxiliary side communication, providing a solution for robust wireless power supply to complex loads under dynamic conditions. By introducing the relay coil, constant frequency operation is achieved even when the coupling coefficient changes, relaxing the need to maintain equal coupling coefficients among the three coils. This provides a broader design space for power electronic devices, significantly simplifying the system design challenges, and reducing stringent constraints on system structure and distance. Moreover, the output reference voltage is obtained by collecting the current on the transmitter side without the information on the receiver side. The closed-loop voltage feedback system is employed to control the output voltage of the front-end buck circuit, ensuring constant power output in the strong-coupling region when the load changes.

\section{Theory and Methods} \label{sec:theory-and-methods}
\subsection{Eigen-mode analysis within the coupled mode theory} \label{subsec:modeling-via-coupled-mode-theory}
\begin{figure}[!ht]
  \centering
  \includegraphics[width=3.3in]{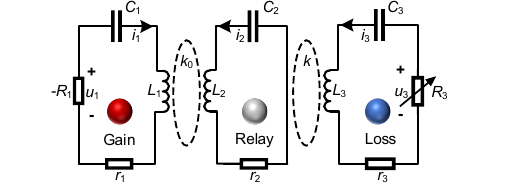}
  \caption{\textbf{Schematic of a third-order pseudo-Hermitian WPT with relay coil.} The red and blue units represent the gain and loss, respectively, which consist of $-$RLC and RLC resonators. The gray unit represents neutral, which consists of the LC resonator. The parasitic resistance of the coils is indicated by $r_n$ ($n = 1, 2, 3$). \label{fig:figure01}}
\end{figure}

We begin with a third-order WPT system, which consists of coupled resonators of a source, a relay, and a load, denoted by $n = 1, 2, 3$, respectively, as shown in \figref{fig:figure01}. Here, the load (gain) is comprised of a ($-$)RLC resonator, and the relay consists of an LC resonator. When the coupling coefficient between the source and load is not considered, the system equation regarding the current phasors $\dot{I}_n$ can be obtained according to the Kirchhoff's voltage law, which reads
\begin{equation} \label{eq:circuit-theory-matrix-star}
  \begin{bmatrix}
    \dfrac{-R_1 + r_1}{L_1} + \text{i}\dfrac{X_1}{L_1} & \text{i}\omega k_0\sqrt{\dfrac{L_2}{L_1}} & 0 \\
    \text{i}\omega \sqrt{\dfrac{L_1}{L_2}}k_0 &\dfrac{ r_2}{L_2} + \text{i}\dfrac{X_2}{L_2} & \text{i}\omega \sqrt{\dfrac{L_3}{L_2}}k \\
    0 & \text{i}\omega \sqrt{\dfrac{L_2}{L_3}}k & \dfrac{R_3 + r_3}{L_3} + \text{i}\dfrac{X_3}{L_3}
  \end{bmatrix}
  \begin{bmatrix}
    \dot{I}_1 \\
    \dot{I}_2 \\
    \dot{I}_3
  \end{bmatrix} = 0,
\end{equation}
where $\omega$ is the operating angular frequency; $X_n = \omega L_n - 1/(\omega C_n)$ denotes the impedance of the $n$-th resonator, with $L_n$ and $C_n$ respectively being the coil inductance and tuning capacitance; $r_n$ denotes the parasitic resistance of the coils. Here, the coupling coefficients are defined as $k_0 = k_{12} = M_{12}/\sqrt{L_1 L_2}$ and $k = k_{23} = M_{23}/\sqrt{L_2 L_3}$, where $M_{12}$ is the mutual inductance between the gain and the relay and $M_{23}$ is the mutual inductance between the relay and the load. 

We consider a specific scenario where the natural angular frequencies of the three resonant circuits are identical and equal to $\omega_0$. Furthermore, we assume that the inductance and capacitance in each resonator are, respectively, all equal, i.e. $L_1 = L_2 = L_3 = L$, $C_1 = C_2 = C_3 = C$, and $r_1 = r_2 = r_3 = r$. As a result, the system equation \eqref{eq:circuit-theory-matrix-star} can be rewritten as
\begin{equation} \label{eq:circuit-theory-matrix}
  \begin{bmatrix}
    -g + \text{i}\bomega & \text{i}\womega k_0 & 0 \\
    \text{i}\womega k_0 & \gamma_\text{r}+\text{i}\bomega & \text{i}\womega k \\
    0 & \text{i}\womega k & \gamma_\text{l} + \text{i}\bomega
  \end{bmatrix}
  \begin{bmatrix}
    \dot{I}_1 \\
    \dot{I}_2 \\
    \dot{I}_3
  \end{bmatrix} = 0,
\end{equation}
where $\bomega = \womega - 1/\womega$ is the normalized angular frequency with $\womega = \omega / \omega_0$; $g = (R_1 - r) \sqrt{C/L}$ denotes the gain parameter of the transmitter,  $\gamma_\text{r}  =  r \sqrt{C/L}$ denotes the loss parameter of the relay and $\gamma_\text{l} = (R_3 + r) \sqrt{C/L}$ denotes the loss parameter of the load. Consequently, non-trivial solutions to \eqref{eq:circuit-theory-matrix} exist only when the determinant of the coefficient matrix is zero, which yields the characteristic equation reading as
\begin{equation} \label{eq:coefficient-matrix-circuit-theory}
    c_6\womega^6 + c_5\womega^5 + c_4\womega^4 + c_3\womega^3 + c_2\womega^2 + c_1\womega + 1 = 0,
\end{equation}
where the coefficients read $c_6 = k^2 + k_0^2 - 1$, $c_5 = -\text{i} (g - g k^2 - \gamma + k_0^2 \gamma_\text{l} - \gamma_\text{r})$, $c_4 = 3 - k^2 - k_0^2 - g (\gamma_\text{l} + \gamma_\text{r}) + \gamma_\text{l} \gamma_\text{r}$, $c_3 =\text{i} (2g - 2\gamma_\text{l} - 2\gamma_\text{r} + g \gamma_\text{l} \gamma_\text{r})$, $c_2 = g (\gamma_\text{l} + \gamma_\text{r}) - \gamma_\text{l} \gamma_\text{r} - 3$, and $c_1 = \text{i} (\gamma_\text{l} + \gamma_\text{r} - g)$. In general, it is not possible to obtain analytical expressions of solutions to the sixth-order characteristic equation \eqref{eq:coefficient-matrix-circuit-theory}. Fortunately, in the weak coupling regime when $k_0^2 \ll 1 $ and $k^2 \ll 1$ and when the operational frequency closely matches the natural resonant frequency, it is possible to apply the coupling mode theory (CMT), which would simplify the analysis.

Within the CMT, the system equation \eqref{eq:circuit-theory-matrix} can be expressed in terms of the energy $a_n = \sqrt{L_n/2}I_n \propto e^{\text{i}\omega t}$ in the $n$-th resonators, which reads $\text{i}{\text{d} \vect{a}}/{\text{d}t} = \vect{H}_\text{pse} \vect{a}$ in the time domain considering $\text{i}\omega \to \text{d}/\text{d}t$ according to the Fourier transform, where $\vect{a} = [a_1, a_2, a_3]^\text{T}$. Here, the Hamiltonian $\vect{H}_\text{pse}$ reads
\begin{equation} \label{eq:pseudo-Hermitian-Hamiltonian}
  \vect{H}_\text{pse} = \frac{\omega_0}{2}
  \begin{bmatrix}
    2 + \text{i}g & k_0 & 0 \\
    k_0 & 2 - \text{i}\gamma_\text{r} & k \\
    0 & k & 2 - \text{i}\gamma_\text{l}
  \end{bmatrix}.
\end{equation}
Note that we have neglected the cross-coupling between the transmitter and the receiver. In our analysis, all resonators are set to resonate at the natural resonance frequency $\omega_0$. 

Now, according to $\text{det}(\omega\vect{I} - \vect{H}_\text{pse}) = 0$, where $\vect{I}$ denotes an identity matrix, one can obtain the characteristic equation by separating the real and imaginary parts, which read
\begin{subequations} \label{eq:characteristic-equation}
\begin{align}
    (\womega-1) \left[ k^2 + k_0^2 - g (\gamma_\text{l} + \gamma_\text{r} )+ \gamma_\text{l}\gamma_\text{r} - 4 (\womega-1)^2 \right] &= 0, \\
    \text{i} \left[ (\womega-1)^2 (g-\gamma_\text{l}-\gamma_\text{r}) + \frac{1}{4} \left(\gamma_\text{l} k_0^2 - g k^2 - g \gamma_\text{l}\gamma_\text{r} \right) \right] &= 0.
\end{align}
\end{subequations}
In practical applications, the loss parameter $\gamma_\text{l}$ and $\gamma_\text{r}$ is generally known; also, the coupling coefficient $k_0$ between the transmitter and the relay coils is given. Therefore, it is necessary to solve for the corresponding gain parameter $g$ and the eigenfrequency $\womega$ when the coupling between the load and the relay changes, namely the parameter $k$. Consequently, the eigenfrequencies can be derived as
\begin{subequations} \label{eq:operating-frequencies}
\begin{align}
    \womega_1 &= 1, \label{eq:operating-frequencies-1} \\
    \womega_{2,3} &= 1 \pm \frac{1}{2} \sqrt{k^2 + k_0^2 + \frac{1}{2} \left( -k_0^2 - \gamma_\text{l}^2 - \gamma_\text{r}^2 + \sqrt{c} \right)}, \label{eq:operating-frequencies-2} \\
    \womega_{4,5} &= 1 \pm \frac{1}{2} \sqrt{k^2 + k_0^2 + \frac{1}{2} \left( -k_0^2 - \gamma_\text{l}^2 - \gamma_\text{r}^2 - \sqrt{c} \right)},
\end{align}
\end{subequations}
with the corresponding gain reading
\begin{subequations} \label{eq:operating-gains}
\begin{align}
    g_1 &= \frac{\gamma_\text{l} k_0^2 } {k^2 + \gamma_\text{l} \gamma_\text{r}}, \label{eq:operating-gains-1} \\
    g_{2,3} &= \frac{1}{2 (\gamma_\text{l} + \gamma_\text{r})} \left[ k_0^2 + (\gamma_\text{l} + \gamma_\text{r})^2 \mp \sqrt{c} \right], \label{eq:operating-gains-2} 
\end{align}
\end{subequations}
where $c = - 4 k^2 (\gamma_\text{l} + \gamma_\text{r})^2 + \left( k_0^2 + \gamma_\text{l}^2 - \gamma_\text{r}^2 \right)^2$. We note that when $k = k_0$ and relay coil loss  $\gamma_\text{r} = 0$, the system becomes a standard third-order PT symmetry circuit, where $g$ is always equal to $\gamma_\text{l}$, but the requirement of coupling coefficients limits the practicality of the system. 

It is evident that the coupling coefficients, the gain and load parameters all affect the operating states, i.e. the eigenfrequencies \eqref{eq:operating-frequencies} and the corresponding eigenmodes. The system can remain stable provided that one of the eigenfrequencies \eqref{eq:operating-frequencies} is real. Remarkably, when the gain resonator can be tuned to satisfy \eqref{eq:operating-gains-1} when $k$ varies for given $\gamma_\text{l}$, $\gamma_\text{r}$ and $k_0$, a stable state corresponding to the stable frequency $\womega = 1$ always exists, as indicated by \eqref{eq:operating-frequencies-1}.

\begin{figure}[!ht]
  \centering
  \includegraphics[width=3.3in]{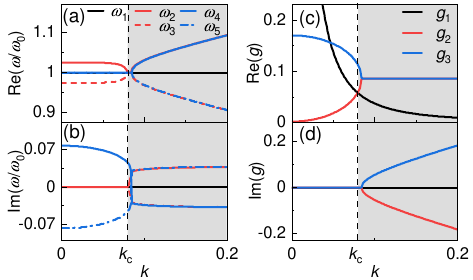}
  \caption{\textbf{Evolution of the eigenmodes as the coupling coefficient $k$ varies within the CMT.} (a) Real and (b) imaginary parts of the eigenfrequencies with respect to the coupling coefficient $k$. (c) Real and (d) imaginary parts of the gain with respect to the coupling coefficient $k$. Here, $k_\text{c}$ indicates the critical coupling coefficient when $\omega$ (or $g$) bifurcates. \label{fig:figure02}}
\end{figure}

\figref{fig:figure02} plots the evolution of the theoretical eigenfrequencies and gain with respect to the coupling coefficient $k$, when $\gamma_\text{l} = 0.154$, $\gamma_\text{r} = 0.002$ and $k_0 = 0.05$. We notice that $\omega_1$ always remains a real number, indicating that the corresponding mode is a possible operating state. When $k \ge k_\text{c}$, both $\omega_{2,3}$ and $\omega_{4,5}$ are complex numbers, indicating that the system can only operate in the $\omega_1$ mode. When $k < k_\text{c}$, $\omega_{2,3}$ are real numbers, and $\omega_{4,5}$ are complex numbers, implying that the system cannot operate in the $\omega_{4,5}$ mode. Since the system can operate in only one steady state at a given time, it is essential to clarify whether the system operates at $\omega_1$ or $\omega_{2,3}$. According to the principle of minimum energy, the system tends to stabilize in the state of minimum energy. In other words, if there are multiple steady states with different corresponding gains, the circuit will tend to operate with the smallest gain. When $k < k_\text{c}$, the gain $g_2$ corresponding to $\omega_{2,3}$ is smaller than the gain $g_1$ corresponding to $\omega_1$, which leads the system to operate in the $\omega_{2,3}$ mode. 

\begin{figure}[!ht]
  \centering
  \includegraphics[width=3.3in]{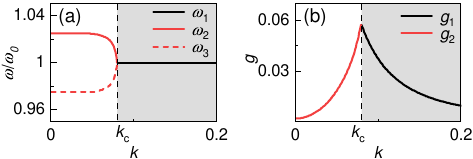}
  \caption{\textbf{Eigenmode and gain selections mechanism for various coupling coefficients.} (a) Evolution of the actual eigenmodes with respect to the coupling coefficient $k$ and (b) the corresponding gains.\label{fig:figure03}}
\end{figure}

\figref{fig:figure03} plots the actual eigenfrequencies and gains with respect to the coupling coefficient $k$. Here, all eigenfrequencies and gain are purely real. The critical coupling coefficient $k_\text{c}$ determines the range of frequency transitions, i.e. when $k \ge k_\text{c}$, the system can achieve constant frequency operation. By solving $\Re[g_2] = g_1$ and $\Re[g_3] = g_1$, we can obtain the critical coupling coefficients $k_\text{c}$ reading as
\begin{subequations} \label{eq:kc}
\begin{align}
    k_\text{c1} & = \sqrt{ \frac{ \gamma_\text{l} \left(k_0^2 - \gamma_\text{l} \gamma_\text{r} - \gamma_\text{r}^2 \right)}{\gamma_\text{l} + \gamma_\text{r}}}, \label{eq:kc_1} \\
    k_\text{c2} & = \sqrt{\frac{1}{2} \left( k_0 \sqrt{k_0^2 + 4\gamma_\text{l}^2 +4\gamma_\text{l}\gamma_\text{r}} - k_0^2 -2\gamma_\text{l}\gamma_\text{r} \right)}, \label{eq:kc_2}
\end{align}
\end{subequations}
which correspond to the two intersections of the curves $g-k$ in \figref{fig:figure02}{(c)}, respectively. When the parameters change, it is necessary to compare the magnitudes of the gains to determine $k_\text{c}$. Combining the evolution of the gain curves in \figref{fig:figure02}{(c)}, it is evident that the eigenfrequency jump often occurs at larger values of $k$. The key to this change lies in the influence of coupling on the system, emphasizing the complexity of the dynamic behavior of the system and how adjusting the coupling coefficient under different parameter values can achieve stability and desired performance. By analyzing the gain under various parameter settings, we can identify the critical coupling coefficient $k_\text{c}$ and gain a deeper understanding of the characteristics of negative resistance.

\begin{figure}[!ht]
  \centering
  \includegraphics[width=3.3in]{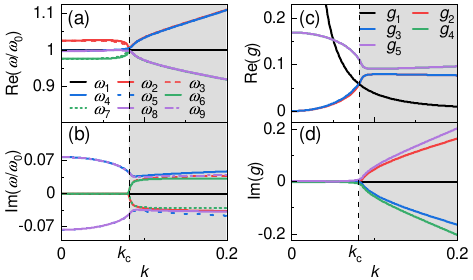}
  \caption{\textbf{Theoretical solutions of modes within circuit theory.} (a) Real and (b) imaginary parts of the eigenfrequencies with respect to the coupling coefficient $k$. (c) Real and (d) imaginary parts of the gain with respect to $k$. Here, $g_n$ corresponds to $\womega_n$ ($n = 1, 2, ..., 6$).\label{fig:figure04}}
\end{figure}

To validate the correctness of the analytical solutions under the CMT, \figref{fig:figure04} presents the numerical solutions of eigenfrequencies and gain with respect to the coupling coefficient $k$, which are obtained using a more strict circuit theory under the same parameters. Although circuit theory solutions indicate that the system has nine eigenfrequencies corresponding to five gains, the eigenfrequencies chosen by the real system under different coupling coefficients $k$ are still consistent with those shown in \figref{fig:figure03}. Therefore, CMT is applicable to analyze the frequency characteristics of the system, especially in the region $k \ge k_\text{c}$, where the system achieves constant frequency stability under dynamic conditions.

\subsection{Analysis of Transmission Characteristics} \label{sec:Analysis-of-Transmission-Characteristics}
After determining the operational mode of the system, we can analyze the transmission characteristics of the WPT system based on pseudo-Hermitian properties ($k \neq k_0$), such as transfer efficiency $\eta$ and output power $P$. In this paper, we consider all the resonant circuits in the system to be series (S) compensation topologies, since such a topology is more suitable for achieving higher power output under low load resistance conditions (less than \si{\kilo\Omega} levels).

In general, the transfer efficiency of a series-compensated WPT system can be defined as
\begin{equation} \label{eq:et_definition}
    \eta = \frac{I_3^2 R_3}{I_1^2 r_1 + I_2^2 r_2 + I_3^2 r_3 + I_3^2 R_3} = \frac{R_3}{r {I_1^2}/{I_3^2} + r {I_2^2}/{I_3^2} + r + R_3},
\end{equation}
where $I_1$, $I_2$ and $I_3$ are the root-mean-square (RMS) values of $\dot{I}_1$, $\dot{I}_2$ and $\dot{I}_3$, respectively; $r_1 = r_2 = r_3 = r$ are the loss of the source and load $LC$ resonators; $R_3$ is the load resistance. Currents $I_1$, $I_2$, and $I_3$ are related when the system operates in pseudo-Hermitian modes, which can be obtained by solving the eigenmodes of \eqref{eq:circuit-theory-matrix}; therefore, the current ratios can be respectively expressed as
\begin{subequations} \label{eq:current-ratio}
\begin{align}
    \frac{\dot{I}_1}{\dot{I}_3} &= \frac{k_0 \left( \gamma_\text{l} + \text{i}\bomega \right)}{k\left( \text{i}\bomega - g \right)},\label{eq:I3/I1} \\
    \frac{\dot{I}_2}{\dot{I}_3} &= -\frac{\gamma_\text{l} + \text{i}\bomega}{\text{i} k \womega}, \label{eq:I3/I2}
\end{align}
\end{subequations}
where $\bomega = \womega - 1/\womega$ is the normalized operating frequency and $g$ is the corresponding gain, which depends on the coupling coefficients $k$ and $k_0$, and the loss parameters $\gamma_\text{l}$ and $\gamma_\text{r}$. As indicated in \figref{fig:figure04} above, since the eigenfrequencies solved within the CMT agree with the actual mode of circuit analysis, the solutions of $\womega$'s and $g$'s in \eqref{eq:operating-frequencies} and \eqref{eq:operating-gains}, can be adopted. In the weak and strong coupling region when $k < k_\text{c}$ and $k > k_\text{c}$, respectively, the system operates in the modes $\womega_{2,3}$ and $\womega = 1$ with the corresponding gains being $g_2$ and $g_1$. Therefore, by substituting \eqref{eq:operating-frequencies}, \eqref{eq:operating-gains} and \eqref{eq:current-ratio} into \eqref{eq:et_definition}, the efficiency $\eta$ of the system can be expressed as
\begin{equation} \label{eq:et}
  \eta(R_3, k) = 
  \begin{cases}
    \cfrac{R_3}{\cfrac{rk_0^2(\gamma_\text{l}^2+\overline{\omega}_{2,3}^2)}{k^2(\overline{\omega}_{2,3}^2+g^2)}+\cfrac{r(\gamma_\text{l}^2+\overline{\omega}_{2,3}^2)}{k^2\widetilde{\omega}_{2,3}^2}+r+R_3}, &  k < k_\text{c}, \\[2.5em]
    \cfrac{k_0^2k^2R_3}{\left( k^2 +\gamma_\text{l}\gamma_\text{r} \right)^2 r +k_0^2\left(\gamma_\text{l}^2 r +k^2r +k^2R_3\right)}, &  k > k_\text{c},
  \end{cases}
\end{equation}
where $\bomega_{2,3} = \womega_{2,3} - 1/\womega_{2,3}$ with $\womega_{2,3} = \womega_{2,3}(k)$ given in \eqref{eq:operating-frequencies-2} since $k_0$ and $\gamma_\text{l}$,  $\gamma_\text{r}$ are known in practical applications.

The output power $P = I_3^2 R_3$ on the load side can be determined by the information on the source side since the currents are related according to \eqref{eq:I3/I1}, which reads
\begin{align} \label{eq:P}
    P = P(R_3, k) = 
    \begin{cases}
        \cfrac{k^2 U_1^2 R_3 \left(\bomega_{2,3}^2 + g_2^2 \right)}{k_0^2 \left( \gamma_\text{l}^2 + \bomega_{2,3}^2 \right) R_1^2}, & k < k_\text{c}, \\[1.5em]
        \cfrac{k_0^2 k^2 U_1^2 R_3}{(k^2 + \gamma_\text{l} \gamma_\text{r})^2 R_1^2}, & k \ge k_\text{c} ,
    \end{cases}
\end{align}
where $R_1$ and $U_1$ are negative resistance and the voltage across the negative resistor, respectively, such that $U_1 = I_1 R_1$ with $R_1 = g_1 \sqrt{L/C} + r = \gamma_\text{l} k_0^2/ (k^2 + \gamma_\text{l}\gamma_\text{r}) \sqrt{L/C} + r$. 

\begin{figure}[!ht]
  \centering
  \includegraphics[width=3.2in]{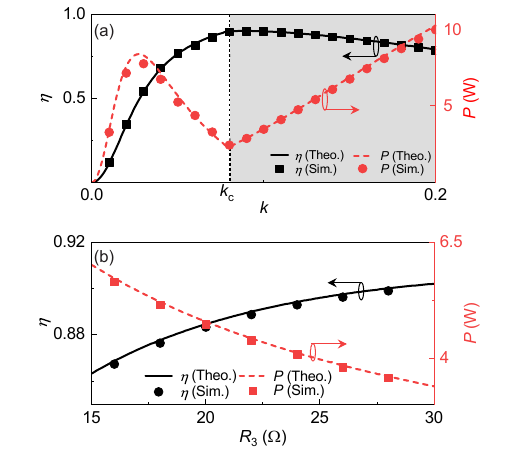} 
  \caption{\textbf{Theoretical (curves) and simulated (markers) output characteristics with respect to (a) the coupling coefficients and (b) the loads.} The solid (black) and dashed (red) curves (markers) show the evolution of transfer efficiency and output power as a function of the coupling coefficient and loads, which are obtained from \eqref{eq:et} and \eqref{eq:P}, respectively. In the analysis, the parameters are the same for both (a) and (b).}\label{fig:figure05}
\end{figure}

\figref{fig:figure05}{(a)} illustrates the theoretical and simulated evolution of the output power $P$ and the efficiency $\eta$ with respect to the coupling coefficient $k$ for a given load resistance $R_3$. In the simulation within PSIM (a software for power electronics and motor drive design and simulation), the parameters are: $U_1 = 4.5~\si{\volt}$, $L = 105~\si{\micro\henry}$, $C = 6.03~\si{nF}$, $R_3 = 20~\si{\ohm}$, $r = 0.3~\si{\ohm}$. When $k < k_\text{c}$ in the weak coupling region, the transfer efficiency of the system increases gradually as the coupling coefficient increases, while the overall efficiency is low; the output power increases first and dramatically decreases. Similar to the traditional second-order PT-WPT system, the output characteristics are unstable in the weak coupling region \cite{assawaworrarit2017robust}. When $k \ge k_\text{c}$ in the strong coupling region, the system has an almost stable transfer efficiency over 80\% despite the fact that the efficiency slightly decreases as the coupling coefficient increases; accordingly, the enhancement of the coupling coefficient would lead to an increase in output power.

\figref{fig:figure05}{(b)} illustrates the efficiency and output power with respect to load resistance when the system operates in the strong coupling region with fixed $k = 0.12$. Here, we have $\sqrt{ \left( k_0 \sqrt{k_0^2 + 4\gamma_\text{l}^2 +4\gamma_\text{l}\gamma_\text{r}} - k_0^2 -2\gamma_\text{l}\gamma_\text{r}\right)/2} < k = 0.12$ when $15~\si{\ohm} < R_3 < 30~\si{\ohm}$ (note $\gamma_{l} = (R_3 + r) / \sqrt{L/C}$), ensuring that the system is always operating at $\omega_0$. For the specific example, the power $P$ received by the load decreases from $6.01~\si{W}$ to $3.39~\si{W}$ when the load resistance $R_3$ increased from $15~\si{\ohm}$ to $30~\si{\ohm}$, indicating a reduction in approximately 45\% power output when the load resistance is doubled. The power fluctuation becomes more noticeable at higher voltage levels. Since the efficiency of the system is relatively high, it is possible to achieve efficient and stable performance by implementing regulation mechanisms when the load fluctuates.

\subsection{Control Strategies for Constant Power Output}
\begin{figure}[!ht]
  \centering
  \includegraphics[width=3.3in]{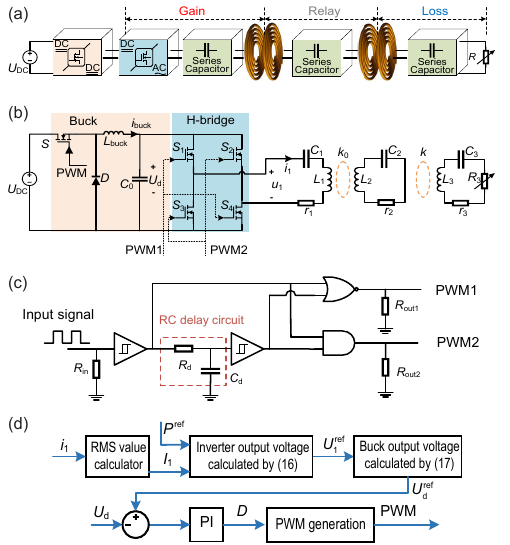}
  \caption{\textbf{Main architecture of a WPT system with stable power output.} (a) Structural diagram and (b) corresponding circuit components of each module. (c) Schematic diagram of dead zone generation. (d) Control block diagram for the buck converter based on voltage closed loop regulation.}
  \label{fig:figure06}
\end{figure}

Now, we discuss the control strategies to achieve constant power output when the load changes. \figref{fig:figure06} illustrates the main architecture and circuit schematic. A full-bridge inverter serves as a nonlinear gain element rather than an operational amplifier to effectively regulate the voltage across the negative resistor and to increase the output power. For the given reference directions shown in \figref{fig:figure06}{(b)}, provided that $u_1$ and $i_1$ on the transmitter side are in phase, the source can be regarded as a negative resistor. Since the system operates at the intrinsic resonant frequency, only a switching signal with a given frequency is required, which benefits the design and control of the buck and H-bridge circuits. Furthermore, to avoid simultaneous conduction of the upper and lower arms of the inverter, it is necessary to add dead time to the control signal and generate pulse width modulation (PWM) signals PWM1 and PWM2 to drive the switch S$_\text{14}$ and S$_\text{23}$, respectively, as shown in \figref{fig:figure06}{(c)}. 

Although the output voltage $u_1$ is a square wave, the output current $i_1$  contains almost the fundamental wave due to the high intrinsic quality factor of the resonator. Therefore, the fundamental RMS value of the inverter output voltage is
\begin{equation} \label{eq:U1-Ud}
    U_1 = \frac{2\sqrt{2}}{\pi} U_\text{d},
\end{equation}
where $U_\text{d}$ is the input voltage of the inverter or the output voltage of the front-end DC-DC converter. Here, we consider a buck converter, whose output voltage reads $U_\text{d} = U_\text{DC} D$, where $U_\text{DC}$ is the voltage of the power supply and $D$ is the duty cycle of the buck circuit.

In practical implementation, it would be much better if a stable output could be realized without communicating between the receiver and transmitter sides, which relaxes the design of the communication unit and minimizes the device size. We aim to detect changes in load by sampling transmitter side parameters such as $u_1$ and $i_1$ when the load changes. Interestingly, the characteristics of the pseudo-Hermitian circuit reveal that $R_3 = \gamma_\text{l} \sqrt{L/C} - r$ is related to the negative resistance $R_1 = g_1\sqrt{L/C} + r = U_1 / I_1$, and $g_1 = \gamma_\text{l} k_0^2/(k^2 + \gamma_\text{l}\gamma_\text{r})$ in the strong coupling region at the stable-frequency mode $\womega = 1$; thus, the load resistance $R_3$ can be re-written in terms of the voltage $U_1$ and current $I_1$ on the transmitting side, which reads
\begin{equation} \label{eq:R1-R3}
    R_3 = \cfrac{ (k^2 + \gamma_\text{r}^2) U_1 - (k^2 + k_0^2 + \gamma_\text{r}^2) I_1 r} {I_1 (k_0^2 + \gamma_\text{r}^2) r - \gamma_\text{r}^2 U_1} r,
\end{equation}
where $\gamma_\text{r} = r\sqrt{C/L}$ denotes the loss coefficient for the parasitic resistance of the coils. Thus, considering \eqref{eq:I3/I1}, the output power $P = I_3^2 R_3$ in \eqref{eq:P} in the strong coupling region when $k > k_\text{c}$ can be expressed in terms of the voltage $U_1$ and current $I_1$ on the source side, which reads as 
\begin{equation} \label{eq:U-P_I}
\begin{split}
    P = \cfrac{ \left[I_1 r (k_0^2 +  \gamma_\text{r}^2) - \gamma_\text{r}^2 U_1 \right] \left[(k^2 + \gamma_\text{r}^2) U_1 - I_1 r (k^2 + k_0^2 + \gamma_\text{r}^2) \right]}{r k^2 k_0^2}.
\end{split}
\end{equation}
Note that when the coils are lossless, i.e., $r = 0$, we will have $P = U_1 I_1$; that is, the transmitting efficiency will become 100\%. Now, once the positions of the coils are fixed (that is, $k$ and $k_0$ are known) and the parasitic resistance $r$ of the coils is known, it is possible to deliver a constant output power received by the load only according to the RMS values of the voltage and current at the transmitter end without the load information, which would significantly benefit practical applications. Therefore, given an output power of $P^\text{ref}$, the reference value of the voltage across the negative resistor is
\begin{equation} \label{eq:U1ref}
    U_1^\text{ref} = \frac{P^\text{ref}}{I_1} \cfrac{2}{1 + \sqrt{1 - \cfrac{4P^\text{ref}}{I_1^2 r} \cfrac{\gamma_\text{r}^2(k^2 + \gamma_\text{r}^2)}{k^2 k_0^2}}} + I_1 r \frac{k^2 + k_0^2 + \gamma_\text{r}^2}{k^2 + \gamma_\text{r}^2};
\end{equation}
and the reference output voltage of the converter reads 
\begin{equation} \label{eq:Udref}
    U_\text{d}^\text{ref} = \sqrt{2} \pi U_1^\text{ref} / 4
\end{equation}
according to \eqref{eq:U1-Ud}, which is used to generate the PWM control signals for the switch in the buck converter.

To achieve constant power output, a closed-loop proportional integral (PI) control method is proposed, and the control block diagram is shown in \figref{fig:figure06}{(d)}. When the load $R_3$ changes, the fundamental RMS value of the current $I_1$ is obtained according to the output current of the inverter $i_1$; accordingly, the reference value $U_\text{d}^\text{ref}$ of the output voltage of the buck circuit is calculated using \eqref{eq:U1ref}. We stress that an amplitude limiter should be included to avoid negative values at the root in \eqref{eq:U1ref} when the current $I_1$ is small. Consequently, the deviation between the reference voltage and the real output voltage serves as the input to the PI voltage regulation loop, which determines the duty cycle $D$ and produces the PWM signal required to drive the switch in the buck converter.

\section{Results and Discussion}
\begin{figure*}[!ht]
    \centering
    \includegraphics[width=6.9in]{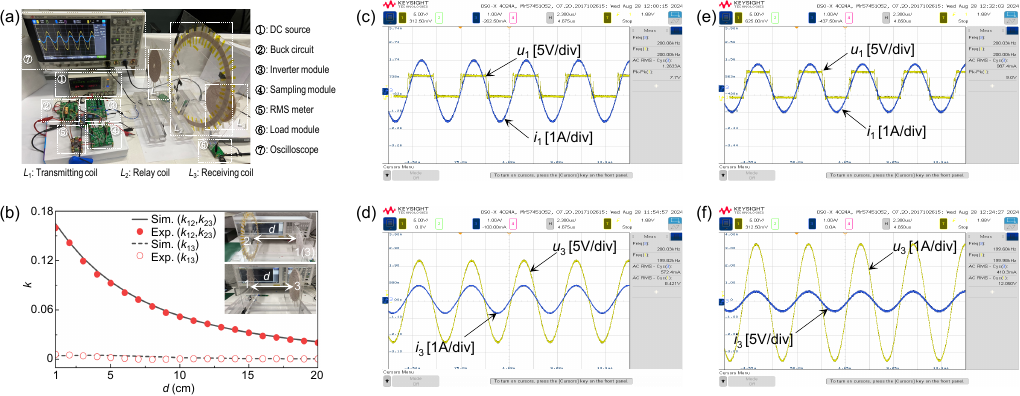}
    \caption{\textbf{Experimental setup and measured voltage and current waveforms on the transmitting and receiving sides.} (a) Photo of the experimental setup. (b) Measured and simulated coupling coefficients at different separation distances. The solid black curve (red dots) corresponds to the coupling coefficient $k_{12}$($k_{23}$) between the relay and transmitting (receiving) coils; the gray dashed curve (circles) corresponds to $k_{13}$. In the experiment, coils 1 and 3 are deliberately misaligned so that $k_{13}$ is small and negligible. Snapshots of the voltage (yellow) and current (blue) waveforms on the (c,e) source and (d,f) load sides when the coupling coefficient is $k = 0.1$. The columns in the middle (c,d) and in the right (e,f) correspond to $R_3 = 15~\si{\Omega}$ and $R_3 = 30~\si{\Omega}$, respectively. } \label{fig:figure07}
\end{figure*}

\figref{fig:figure07}{(a)} illustrates the experimental prototype. Starting from the main power circuit, the input DC voltage passes through the front-end buck converter, which comprises an N-channel MOSFET IRFP4322PBF, a Schottky diode MBRF2545CTG, and an isolated gate driver chip Si8271. The DC voltage $U_\text{d}$ undergoes a high frequency inversion through four MOSFET BSC160N10NS3G chips, which transform into a square wave voltage at a frequency of 200~kHz to energize the transmitter. In the primary-side control circuit, the current of the LC circuit is sensed by current transformer CU8965. The voltage and current signals are processed through a differential amplification circuit consisting of the operational amplifier OPA2690 and the RMS conversion chip AD637. Various PWM signals are generated by the STM32F103 microcontroller from STMicroelectronics.

For the transmitting and receiving coils in the experiments, we used resonant coils wound with $0.05~\si{mm} \times 500$ Litz wire to minimize internal resistance. To minimize the volume of occupied space, we chose planar spiral coils for the compact design. The self-inductance of planar spiral coils can be determined by evaluating the ratio of the magnetic flux generated by the conductor to the current, which can be approximated using a closed form reading as \cite{mohan1999simple}
\begin{equation} \label{eq:self-inductance}
    L \approx 0.5 \mu_0 N^2 d_\text{av} \left( 0.9 + 0.2f^2 -\text{ln}f \right),
\end{equation}
where $\mu_0$ is the vacuum permeability; $N$ is the number of turns; $d_\text{av} = (d_\text{out} + d_\text{in})/2$ is the average diameter of the coil with $d_\text{out}$ and $d_\text{in}$ denoting outer and inner diameters, respectively; $f = (d_\text{out} + d_\text{in})/(d_\text{out} - d_\text{in})$ is filling rate. Here, $d_\text{out} = d_\text{in} + (2 N + 1) w + (2 N - 1) s$ where $w$ and $s$ are the width and spacing of the spiral coil, respectively. In general, because of the presence of parasitic capacitance in the coils, the actual inductance would increase as the frequency increases, which would increase the complexity of the design. However, since we consider the system to operate in a constant-frequency mode, the inductance of the coil remains unchanged at its natural resonant frequency, which would benefit the design.

\figref{fig:figure07}{(b)} illustrates the coupling coefficient $k$ between the relay coil and the transmitting (receiving) coil as the separation distance $d$ increases. The geometric parameters for the transmitter and receiver coils are: $N = 38.9$, $d_\text{in} = 20~\si{mm}$, $s = 0.13~\si{mm}$, and $w = 2.16~\si{mm}$; and the parameters read $N = 10.83$, $d_\text{in} = 350~\si{mm}$, $s = 0.13~\si{mm}$, and $w = 2.16~\si{mm}$ for the relay coils.  The results of the full-wave simulation are consistent with the experimental results, which are measured using a Vector Network Analyzer (VNA). It can be observed that the coupling coefficient $k$ decreases as the separation distance $d$ between the coaxial coils increases. Note that the effective transmission distance generally does not exceed the diameter of the coils. As a reference, the coupling coefficient $k_{13}$ between the transmitting coil and the receiving coil is plotted, which is too small to be neglected since the transmitting and receiving coils are deliberately eccentrically aligned to minimize mutual coupling. In addition, the coupling coefficient between two concentric coils differs from the adjacent coupling of three concentric coils for a given distance. 

\begin{figure}[!ht]
   \centering
    \includegraphics[width=3.3in]{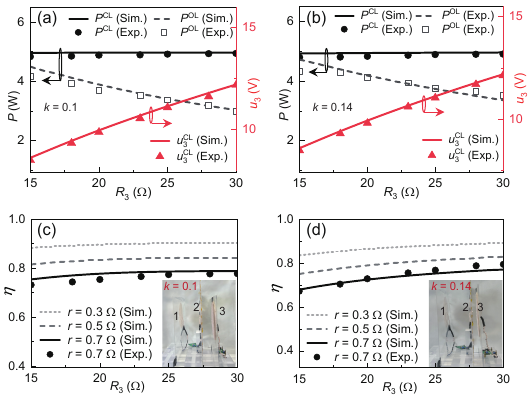}
    \caption{\textbf{Comparison of experimental and simulated results of the output power and the transmission efficiency.} (a) and (b) display the output power and voltage for load when $k = 0.1$ and $k = 0.14$, respectively. The black solid curves (dots) and red solid curves (triangles) represent the power received and the voltage at the load under closed-loop control (CL), respectively; while the received power at the load under open-loop (OL) control is depicted in the gray dashed curves (hollow squares). (c) and (d) show the transmission efficiency when $k = 0.1$ and $k = 0.14$, respectively. As references, the simulated transmission efficiency is depicted in gray (dashed curves), when the parasitic resistance is small, that is, $r = 0.3~\si{\ohm}$ and $r = 0.5~\si{\ohm}$. In the experiments, the reference output power is $P^\text{ref} = 5~\si{W}$.}\label{fig:figure08}
\end{figure}

We implemented a voltage PI loop feedback control strategy to enhance stability and dynamic performance. The parameters of the voltage loop (whose input is $U_D$) are $k_\text{P} = 0.01$ and $k_\text{I}=0.6$. \figref{fig:figure07}{(c)} -- \figref{fig:figure07}{(f)} display the measured voltage and current waveforms in the strongly coupled region when $k = 0.1$ for load $R_3 = 15~\si{\Omega}$ (in the middle column) and $R_3 = 30~\si{\Omega}$ (in the right column), respectively. In the experiments, the reference power is $P^\text{ref} = 5~\si{W}$. When the system operates in a strongly coupled region, the voltage $u_1$ across the inverter and the current $i_1$ at the transmission end are always in phase, showing the negative resistance characteristic in the resonant state. In addition, the system operates at a stable frequency of 200~\si{kHz}. As resistance to the load $R_3$ increases, voltages $u_1$ and $u_3$ increase (also see red curves and markers in \figref{fig:figure08}{(a)} and \figref{fig:figure08}{(b)}), while currents $i_1$ and $i_3$ decrease, ensuring that the power received by the load remains constant under PI control, as shown in \figref{fig:figure07}{(d)} and \figref{fig:figure07}{(f)}.

\figref{fig:figure08} compares the experimental and simulated results of the output power and efficiency characteristics for different coupling coefficients $k_{23}$ when the load $R_3$ changes. Here, the coupling coefficient between the transmitting coil and the relay is $k_{12} = k_0 = 0.05$, and the coupling coefficient between the relay and the receiving coil is $k_{23} = k = 0.1$ and $k_{23} = k = 0.14$, indicating that the system operates in the strongly coupled region for $k > k_c = 0.097$ when the load $R_3$ varies within the range of 15~\si{\Omega} to 30~\si{\Omega} according to \eqref{eq:kc_2}. The results of the experiment are in good agreement with the results of the PSIM simulation, where the parameters are summarized in \tabref{tab:table1}. Within the proposed closed-loop control strategy, the power received by the load is almost constant and follows the reference value of $P^\text{ref} = 5~\si{W}$ with only a maximum relative fluctuation of 2.22\% (2.24\%), compared to the open-loop system that exhibits a maximum relative fluctuation of 32.5\% (28.9\%), as illustrated in \figref{fig:figure08}{(a)}[\figref{fig:figure08}{(b)}]. The maximum error between the experimental value and the reference power is 3.40\% (3.80\%). As shown in \figref{fig:figure08}{(c)} and \figref{fig:figure08}{(d)}, although the transmission efficiency is not as high as conventional \cite{assawaworrarit2020robust}, it can be improved by reducing the parasitic resistance of the coils in the circuit (see the simulated results (gray dashed curves) in \figref{fig:figure08}{(c)} and \figref{fig:figure08}{(d)}) or by introducing frequency-dependent gains \cite{hao2024dispersive}. 
\begin{table}[!ht]
  \caption{Component parameters for PSIM simulations.}\label{tab:table1}
  \begin{tabular}{ccc}
    \hline
    Parameters & Descriptions & Values \\
    \hline
    $U_\text{DC}$ & DC input voltage of the buck converter & $9~\si{\volt}$ \\
    $L_\text{buck}$ & Inductance in the buck converter & $1500~\si{\micro\henry}$ \\ 
    $C_0$ & DC voltage regulator capacitance & $47~\si{\micro F}$ \\
    $C$ & Series capacitance in all three resonators & $6.03~\si{\nano F}$ \\
    $L$ & Inductance in all three resonators & $105~\si{\micro\henry}$ \\
    $r$ & Parasitic resistance of $L$ & $0.7~\si{\ohm}$ \\
    $k_0$ & Coupling coefficient between gain and relay & 0.05 \\
    \hline
  \end{tabular}
\end{table}

\section{Conclusions}
We have proposed a pseudo-Hermitian model for a series-resonant wireless power transfer topology, which exhibits a robust stable frequency and constant power output when the load varies, whilst no direct monitoring of the load condition is required. By introducing a relay coil and forming a third-order pseudo-Hermitian system, it is possible to operate in a constant-frequency mode with high transfer efficiency in the strong-coupling region according to the coupled-mode theory (CMT). The required nonlinear gain can be determined according to the load and coupling coefficients. More importantly, the currents on the source and load sides are related within the given modes, which allows us to monitor only the source side for control purposes, rather than the load, which avoids communication between the transmitting and receiving ends. Furthermore, a control strategy that uses only primary side control is proposed to stabilize output power when the load changes, eliminating the need for end-to-end control circuits and wireless communication. This, in turn, would significantly simplify the control algorithm and reduce the complexity of the circuit. The experimental results have validated the performance of the proposed control strategy and system. Our research provides a reliable solution for constant power wireless charging, particularly in scenarios with dynamic load changes and/or with certain spatial freedom, such as drones and household applications.

\section*{Acknowledgements}
\paragraph*{Funding:} T.D. acknowledges the support of the National Natural Science Foundation of China (51977165) and the Key Research and Development Program of Shaanxi Province (2024GX-YBXM-236). K.Y. thanks the National Natural Science Foundation of China (52407015) and the Postdoctoral Fellowship Program of CPSF (GZB20240469).

\paragraph*{Data and Code Availability:} All data needed to evaluate the conclusions are presented in the article. Raw data, corresponding simulation data, and the relevant code are available upon reasonable request. 

\paragraph*{Competing Interests:} The authors declare that they have no competing interests.

\bibliography{main}

\end{document}